\documentstyle[12pt,aaspp4,psfig]{article}

\newcommand{\kms}          {\mbox{${\rm km~s^{-1}}$}}
\newcommand{\cc}           {\mbox{${\rm cm^{-3}}$}}

\newcommand{\e}            {\mbox{$^{-1}$}}

\newcommand{\eee}          {\mbox{$^{-3}$}}
\newcommand{\simgt}        {\gtrsim}
\newcommand{\simlt}        {\lesssim}

\def\cm2{\mbox{${\rm cm^{-2}}$}}
\def\h2{\mbox{${\rm H}_2$}}
\def\nh2{\mbox{$n_{\rm H_2}$}}
\def\Nh2{\mbox{$N_{{\rm H}_2}$}}
\def\Mh2{\mbox{$M_{{\rm H}_2}$}}
\def\n2hp{\mbox{N$_2$H$^+$}}
\def\c34s{\mbox{C$^{34}$S}}
\def\fs{\hbox{$.\!\!^{\rm s}$}}

\begin{document}

\title{A CONTRACTING, TURBULENT, STARLESS CORE IN THE SERPENS CLUSTER}
\author{Jonathan P. Williams\footnotemark\footnotetext{Current address:
National Radio Astronomy Observatory, Campus Building 65,
949 N. Cherry Avenue, Tucson, AZ 85721-0665; jpwilliams@nrao.edu}
and Philip C. Myers}
\affil{Harvard--Smithsonian Center for Astrophysics, 60 Garden Street,
Cambridge, MA 02138; jpw@cfa.harvard.edu,pmyers@cfa.harvard.edu}

\begin{abstract}
\rightskip = 0pt
We present combined single-dish and interferometric
CS(2--1) and \n2hp(1--0) observations of a compact core in the
NW region of the Serpens molecular cloud.
The core is starless according to observations from optical to millimeter
wavelengths and its lines have turbulent widths and ``infall asymmetry''.
Line profile modeling indicates supersonic inward motions
$v_{\rm in}\simgt 0.34$~\kms\ over an extended region $L>12000$~AU.
The high infall speed and large extent exceeds the predictions of most
thermal ambipolar diffusion models and points to a more dynamical process
for core formation.
A short (dynamic) timescale, $\sim 10^5~{\rm yr}\simeq L/v_{\rm in}$,
is also suggested by the low \n2hp\ abundance $\sim 1\times 10^{-10}$.
\end{abstract}


\rightskip = 0pt
\section{Introduction}
Turbulence is a ubiquitous feature of the interstellar medium
although its precise nature is poorly understood
and its role in the formation of stars is unclear.
Theories of isolated star formation generally assert that
gravitational collapse occurs onto a thermally supported core
(e.g., Shu, Adams, \& Lizano 1987) and motions are
quasi-static until very late times
(Basu \& Mouschovias 1994; Ciolek \& Mouschovias 1995; Li 1998).
Observations of infall at small scales ($\sim 0.01$~pc)
in the isolated starless core, L1544,
in Taurus are not in contradiction with such theories if it is
indeed sufficiently close to forming a star (Williams et al. 1999),
although large scale motions ($\simgt 0.1$~pc)
do appear to require an alternative explanation (Tafalla et al. 1998),
such as turbulent dissipation (Myers \& Lazarian 1998).

The Serpens molecular cloud
(d=310~pc; de Lara, Chavarria-K, \& Lopez-Molina 1991)
has more embedded YSOs and is more turbulent than the Taurus cloud.
In the northwest region of this cloud lies a cluster of Class 0 sources
(Hurt \& Barsony 1996) which has been the subject of numerous studies in
the millimeter and sub-millimeter regime (Casali, Eiroa, \& Duncan 1993;
McMullin et al. 1994; White et al. 1995; Hurt, Barsony, \& Wootten 1996;
Wolf-Chase et al. 1998; Testi \& Sargent 1998).
Since the dense cores around Class 0 sources often show the spectral
signature of inward motion (Mardones et al. 1997), we embarked on a
study of the dense gas dynamics in this young cluster forming region
in order to compare with more isolated star forming sites such as in
Taurus. In this {\it Letter}, we present observations of a previously
unrecognized core adjacent to the Class 0 source S68N that appears to
be starless, contracting, and is highly turbulent. We compare its
properties with its neighbor, S68N,
deduce a chemical timescale for its formation that suggests that it
is very young, and determine an average infall speed by spectral line
modeling. This core, which we designate S68NW, demonstrates that
turbulent motions in the ISM cannot be ignored in the formation of
individual stars in clusters.

\section{Observations}
\label{sec:obs}
A well tested method of diagnosing inward motions onto a star forming
region is to search for self-absorbed lines where emission at low
velocities is brighter than at high velocities
(Leung \& Brown 1977; Walker et al. 1986; Zhou et al. 1993).
We observed CS(2--1) and \n2hp(1--0) toward the cluster of Class 0
sources since both lines are reasonably bright hence
quick to map, both are strongly excited by gas of density
$\nh2\sim 10^5$~\cc,
and generally CS is optically thick and \n2hp\ is optically
thin at the resolution of these observations.

Singledish maps were made at the
Five College Radio Astronomy Observatory\footnotemark
\footnotetext{FCRAO is supported in part by the
National Science Foundation under grant AST9420159 and is operated
with permission of the Metropolitan District Commission, Commonwealth
of Massachusetts} (FCRAO) 14~m telescope in December 1996
using the QUARRY 15 beam array receiver and the FAAS backend
consisting of 15 autocorrelation spectrometers with
1024 channels set to an effective resolution of 24~kHz (0.06~km/s).
The observations were taken in frequency switching mode and, after
folding, 3rd order baselines were subtracted. The pointing and focus
were checked every 3 hours on nearby SiO maser sources.
The FWHM of the telescope beam is $50''$, and a map covering
$6'\times 8'$ was made at Nyquist ($25''$) spacing.

Observations were subsequently made with the 10
antenna Berkeley-Illinois-Maryland array\footnotemark
\footnotetext{Operated by the University of California at Berkeley,
the University of Illinois, and the University of Maryland,
with support from the National Science Foundation}
(BIMA) for two 8 hour tracks in each line during
April 1997 (CS) and October/November 1997 (\n2hp).
A two field mosaic was made with phase center
$\alpha(2000)=18^{\rm h}29^{\rm m}47\fs 5,
~\delta(2000)=01^\circ 15'51\farcs 4$
and a second slightly overlapping pointing at
$\Delta\alpha=33\farcs 0, \Delta\delta=-91\farcs 0$.
Amplitude and phase were calibrated using 4 minute observations of
1751+096 (4.4~Jy) interleaved with each 22 minute integration on source.
The correlator was configured with two sets of 256 channels at
a bandwidth of 12.5~MHz (0.15~\kms\ per channel) in each sideband
and a total continuum bandwidth of 800~MHz. The flexible correlator
setup allowed us to observe CH$_3$OH($2_1-1_1$) in addition to CS(2--1),
and \c34s(2--1) along with \n2hp(1--0): the methanol line was
found to map the outflow associated with S68N (Wolf-Chase et al. 1998)
but the \c34s\ line was detected only marginally.

The data were calibrated and maps produced using standard procedures in
the MIRIAD package. Since the emission is extended, analysis of the spectra
must correct for the spatial filtering properties of the interferometer.
To allow for this, we combined the FCRAO and BIMA data using
the task IMMERGE. Maps were compared within the region of visibility
overlap (6~m to 14~m) and small pointing corrections made to the
FCRAO data ($< 6''$, about a tenth of the beam) which was then scaled
using a gain of 43.7~Jy~K\e.\footnotemark
\footnotetext{Information regarding aperture efficiency measurements
on the FCRAO 14~m telescope can be found on the World Wide Web at
http://donald.phast.umass.edu/$\sim$fcrao/library/techmemos/gain96.html}
The resolution of the resulting maps was
$10\farcs 0\times 7\farcs 8$ at p.a. $-72^\circ$ for CS which was
observed twice in the compact C configuration and
$8\farcs 5\times 4\farcs 6$ at p.a. $+2^\circ$ for \n2hp\ which was
observed once in C configuration and once in the wider B configuration.

\section{Analysis}
\label{sec:analysis}
\subsection{S68N and S68NW}
Analysis of the large scale maps is deferred to a later paper.
Here, we restrict attention to a remarkable region, $\sim 1\farcm 5\times 2'$,
around the S68N protostar. This source was discovered
from earlier 3-element BIMA CS observations by McMullin et al. (1994)
but we re-observed it with the 10-element array to obtain greater
sensitivity and resolution. It was originally undetected in the 3~mm
continuum but is readily apparent in the new data at an integrated
flux level of 12.3~mJy, in agreement with OVRO observations by
Testi \& Sargent (1998).
S68N has also been detected at shorter wavelengths and its spectrum
was fit by a modified blackbody with dust temperature 20~K
and luminosity $5~L_\odot$ by Wolf-Chase et al. (1998).

Maps of the integrated intensity of CS(2--1) and \n2hp(1--0) around S68N
are displayed in Fig.~1. The position of the 3~mm continuum peak,
indicated by the star, lies at the center of the \n2hp\ emission but is
offset by $7"$ from the CS core. However, there is both high velocity
emission from the outflow and self-absorption present in the CS spectra
which may skew the map of integrated intensity relative to the distribution
of dense gas around the star.

Equally striking in the CS map, however, is the presence of a compact core,
hereafter S68NW, that lies $\sim 50''$ west of S68N.
It is also present in the map of \n2hp\ integrated
intensity but is not nearly so prominent. It was not detected in
the continuum to a $3\sigma$ sensitivity of 3.3~mJy~beam\e, nor is it
apparent in the slightly higher sensitivity OVRO Testi \& Sargent observations.
It is also undetectable in maps at
1~mm (Casali et al. 1993; Tafalla \& Mardones, private communication),
at 12, 25, 60, and 100~$\mu$m in the Hurt \& Barsony IRAS HIRES maps,
in the near-infrared ($2~\mu$m; Eiroa \& Casali 1992)
or in the Digital Sky Survey. These observations constrain the
luminosity of any embedded object in S68NW to be less than $0.5~L_\odot$.

\subsection{Abundance differences}
Fig.~1 suggests a difference in the chemistry between
the star forming S68N and starless S68NW core. We have estimated the
abundance of \n2hp\ in the two cores by comparing the mass of \n2hp\
derived from the integrated emission with the virial mass derived from
the size and linewidth.  Given the compact appearance of the cores,
the assumption of virialization is unlikely to be greatly in error.
We define the boundaries of each core as the FWHM contour of the \n2hp\ maps
and calculate sizes, linewidths, and integrated emission within these limits.
Core properties are listed in Table~1.

The inferred virial \n2hp\ abundance is much lower in S68NW than S68N.
This is due to a combination of a smaller size, greater linewidth,
and lower integrated intensity in S68NW, but the relatively low emission
is the dominant factor. In the following section, infall model fits
do not find such an extreme abundance difference between the two
but nevertheless confirms that the
\n2hp\ abundance in S68NW is unusually low, $\sim 1\times 10^{-10}$,
compared to $\sim 4\times 10^{-10}$ in other dense cores
(Womack, Ziurys, \& Wyckoff 1992; Ungerechts et al. 1997).
A potential explanation that
suits the starless nature of S68NW is chemical evolution:
Bergin et al. (1997) show that whereas CS forms very quickly in a
dense core, it takes $\simgt 10^5$~yr to form substantial amounts of \n2hp.
Observations of other time-sensitive molecular species
such as HC$_3$N offer a test of this hypothesis.

\subsection{Spectral line modeling}
The majority of the CS spectra in this region are double-peaked
and the magnitude of the dips between the peaks tends to increase
closer to the core centers.
Average spectra within the FWHM contour of \n2hp\ emission for each core
are displayed in Fig.~2. Unlike the case of L1544 (Williams et al. 1999),
the \n2hp\ spectra are not self-absorbed and we use these data to determine
the velocity and linewidth of the cores.
For each core, the central dip in the CS spectrum lines up with the
\n2hp\ velocity indicating that the CS emission is self-absorbed.
The S68N spectrum shows prominent outflow wings but it is quite symmetric
in marked contrast to S68NW for which the lower velocity
(blue) peak is much brighter than the higher velocity (red) peak.
For a radially decreasing excitation gradient, such as would
exist for a centrally condensed core at constant kinetic temperature,
this indicates that the outer self-absorbing gas is red-shifted (i.e.,
infalling) relative to the inner emitting region.

To estimate the speed of the infalling gas, we have fit the
spectra using a simple two layer model consisting of two
isothermal layers, the near side (to the observer) at low
density and the far side at high density. This model resembles
those discussed by Myers et al. (1996) and Williams et al. (1999):
emission from the rear layer is absorbed by the lower excitation
front layer with the location of the absorption dependent on the
relative velocity between the front and rear layers (i.e., the
infall speed). Observations are used to constrain the models as much
as possible: gaussian fits to the isolated \n2hp\ hyperfine component
are used to set the systemic velocity and linewidth of the core,
the line-of-sight widths of each layer are set equal to the measured
radius of the cores, and the \n2hp\ abundances are constrained to
vary only within a factor of two of the
virial estimates derived in the previous section.
The free parameters are the densities of each layer,
their common kinetic temperature, the molecular abundances,
and the infall speed of the front layer onto the rear layer.
In addition, a low optical depth component between the two layers
was added to the S68N model to allow for the outflow,
and a gaussian component $\sim 2$~\kms\ from line center was
used in the S68NW model to fit excess emission at low velocities.

The model spectra are shown in relation to the observations in Fig.~2
and model parameters listed in Table~2. The rear layer densities are
very similar, approximately equal to the critical density of the
two transitions, and the foreground layer densities are comparable to each
other and similar to the density of $^{13}$CO emitting gas. The kinetic
temperatures are the same for both cores and equal to the dust temperature
of S68N as determined from the spectral energy distribution by
Wolf-Chase et al. (1998). The CS abundances are also
the same and consistent with observations of Orion by Ungerechts et al. (1997)
but the fits require a smaller difference in \n2hp\ abundance than the
virial estimates derived in the previous section (note, however, that the
\n2hp\ abundance of S68NW is still very low).
The parameter that is most different between the two cores is the
infall speed. The low infall speed in S68N
is implied by the near symmetry of the line profile and is little
affected by the addition of the outflow component which is also
quite symmetrical. The inferred infall speed for S68NW, however,
is high because of the large blue-red asymmetry but its precise value
is very sensitive to the strength of an additional gaussian component
added at low velocities. This extra component appears to be physically
unrelated to S68NW; it lacks a red counterpart and peaks in emission
$\sim 2$~\kms\ from the \n2hp\ line. Channel maps suggest that it is
a second CS core, slightly offset along the line of sight.
Its contribution to the integrated CS intensity within one linewidth
of the central velocity of \n2hp\ is less than 20\%
at its peak and is generally much less in other spectra.
The addition of this extra component reduces the blue-red ratio required
in the infall model resulting in a smaller infall speed: in the absence of
this component, the inferred infall speed is $\simgt 0.5$~\kms.
Therefore, we believe that the value listed in Table~2, $0.34$~\kms,
is a {\em lower limit} which implies that the S68NW core is contracting
supersonically ($v_{\rm in}/\sigma_{\rm thermal}({\rm H}_2)>1$)
although not necessarily super-Alfv\'{e}nically
($v_{\rm in}/\sigma_{\rm non-thermal}\simgt 0.5$).
The implied mass infall rate of the front layer onto the
rear layer is $1\times 10^{-7}~M_\odot$~yr\e.

\section{Discussion}
The data presented here indicate that S68NW is a turbulent core
in the process of contraction and increasing its mass substantially.
Furthermore its proximity to the S68N core and
embedded protostar suggests that S68NW may soon form a low-mass star, as
the next part of the sequence of low-mass star formation events which have
occurred in the Serpens complex over the last Myr. If so, the formation of
stars and cores in Serpens may substantially overlap in time, in
contrast to the idea that star formation in clusters is coeval,
such as in response to a single triggering event
(e.g. Zinnecker, McCaughrean \& Wilking 1993). Instead these observations
imply that the cloud is forming a core while its already formed
cores are still forming stars. If so, core formation and star formation
may have relatively similar timescales, each shorter than the overall
cluster formation timescale.

The instantaneous collapse timescale,
$t_{\rm coll}=6200~{\rm AU}/0.34~{\rm km~s^{-1}}\simeq 10^5~{\rm yr}$
is approximately equal to the free-fall time
for gas of density $n({\rm H_2})\sim 10^5$~cm\eee.
Such dynamic motions are achieved in models of ambipolar diffusion
only at very late times, $\sim 10^7$~yr
(Basu \& Mouschovias 1994; Ciolek \& Mouschovias 1995; Li 1998),
which appears to be inconsistent with the low \n2hp\ abundance.
In addition, the supersonic infall speed
requires either low ionization levels, $x_e<10^{-8}$
at $n({\rm H_2})=3\times 10^4$~cm\eee\ (Basu \& Mouschovias 1995a),
significantly less than measured in cores in
Taurus (Williams et al. 1998) and Orion (Bergin et al. 1999),
or a weak magnetic field, $B\simlt 10~\mu$G
at $n({\rm H_2})=5\times 10^3$~cm\eee\ (Basu \& Mouschovias 1995b)
which would imply a smaller Alfv\'{e}n speed, $\sim 0.2$~\kms, than
the observed linewidth. Such a weak field at these relatively high
densities may also conflict with HI Zeeman measurements of similar size
fields at much lower gas densities in Ophiuchus (Goodman \& Heiles 1994).
Finally, ambipolar diffusion models do not predict the large size scales
over which infall occurs: asymmetric, self-absorbed CS line profiles extend
well beyond the FWHM \n2hp\ contour,
indicating detectable inward motions over $\simgt 0.1$~pc.
Similarly large infall zones have also been observed in the isolated core
L1544 (Tafalla et al. 1998) and in the cluster forming regions,
L1251B and NGC1333--IRAS4 (Mardones 1998).

A possible explanation is that the collapse front propagates outward
at the non-thermal, rather than the thermal, sound speed
(e.g. Myers \& Fuller 1993)
in which case the ratio of infall speed to effective sound speed
remains comfortably within the bounds of ambipolar diffusion models.
However, this requires that the non-thermal motions be maintained
in the core over the ambipolar diffusion timescale, $t_{\rm AD}\simeq 10^6$~yr,
but Nakano (1998) shows that the timescale for turbulent dissipation is
approximately the same as the free fall time and much less than $t_{\rm AD}$
in the absence of any internal driving sources.
Indeed, it may be the decay of the non-thermal motions,
with a corresponding loss of pressure support, that drives the fast
inward motions in S68NW (Myers \& Lazarian 1998)
over the large observed size scales.
Although we do not observe a decrease in the \n2hp\ velocity dispersion
toward S68NW, this does not preclude the existence of a small
less turbulent central core.
Observing such an object remains a challenge for the future.
Note that in this scenario, the timescale of the next stage of star
formation would be that of ambipolar diffusion if the core were
magnetically subcritical, or would be dynamical if the core were
magnetically supercritical. It will be useful to determine the incidence
of cores like S68NW in other star-forming regions, and to study such cores
in lines sensitive to a wide range of gas density.

\acknowledgments
This research was partially supported by NASA Origins of Solar Systems
Program, grant NAGW-3401. JPW thanks the Radio Astronomy Laboratory at
the University of California at Berkeley for support during the writing
of the manuscript and Chris McKee and Frank Shu for informative discussions.
Conversations with Shantanu Basu, Glenn Ciolek, and Zhi-Yun Li are also
gratefully acknowledged.



\begin{table}
\begin{center}
TABLE 1\\
\n2hp(1--0) Core Comparison\\
\vskip 2mm
\begin{tabular}{lrr}
\hline\\[-2mm]
                               &  S68N   &  S68NW  \\[2mm]
\hline\hline\\[-3mm]
FWHM Radius (AU)               &  8100   &  6200   \nl
FWHM Linewidth (km s$^{-1}$)   &  0.95   &  1.50   \nl
Virial Mass$^{\rm a}$ ($M_\odot$)
                               &  4.5    &   8.4   \nl
Integrated Intensity (K km s$^{-1}$)
                               &  15.1   &   5.8   \nl
Abundance$^{\rm b}$ ($\times 10^{-10}$)
                               &   5.6   &   0.66  \\[2mm]
\hline\\[-2mm]
\multicolumn{3}{l}{$^{\rm a}$ for an inverse square density profile}\\
\multicolumn{3}{l}{$^{\rm b}$ relative to H$_2$, $T_{\rm ex}=15$~K}\\
\end{tabular}
\end{center}
\label{tab:compare}
\end{table}

\begin{table}
\begin{center}
TABLE 2\\
Infall model parameters\\
\vskip 2mm
\begin{tabular}{lrr}
\hline\\[-2mm]
                               &  S68N   &  S68NW  \\[2mm]
\hline\hline\\[-3mm]
Rear layer density (cm\eee)    &  $2.1\times 10^5$ & $2.2\times 10^5$ \nl
Front layer density (cm\eee)   &  $1.0\times 10^3$ & $2.1\times 10^3$ \nl
Kinetic temperature (K)        &  20.0   &  20.0   \nl
CS abundance$^{\rm a}$ ($\times 10^{-9}$)
                               &   8.0   &   8.0   \nl
N$_2$H$^+$ abundance$^{\rm a}$ ($\times 10^{-10}$)
                               &   2.8   &   1.1   \nl
Infall speed (km s$^{-1}$)     &   0.01  &   0.34  \\[2mm]
\hline\\[-2mm]
\multicolumn{3}{l}{$^{\rm a}$ relative to H$_2$}\\
\end{tabular}
\end{center}
\label{tab:infallfits}
\end{table}
\clearpage

\begin{figure}[htpb]
\vskip -1.3in
\centerline{\psfig{figure=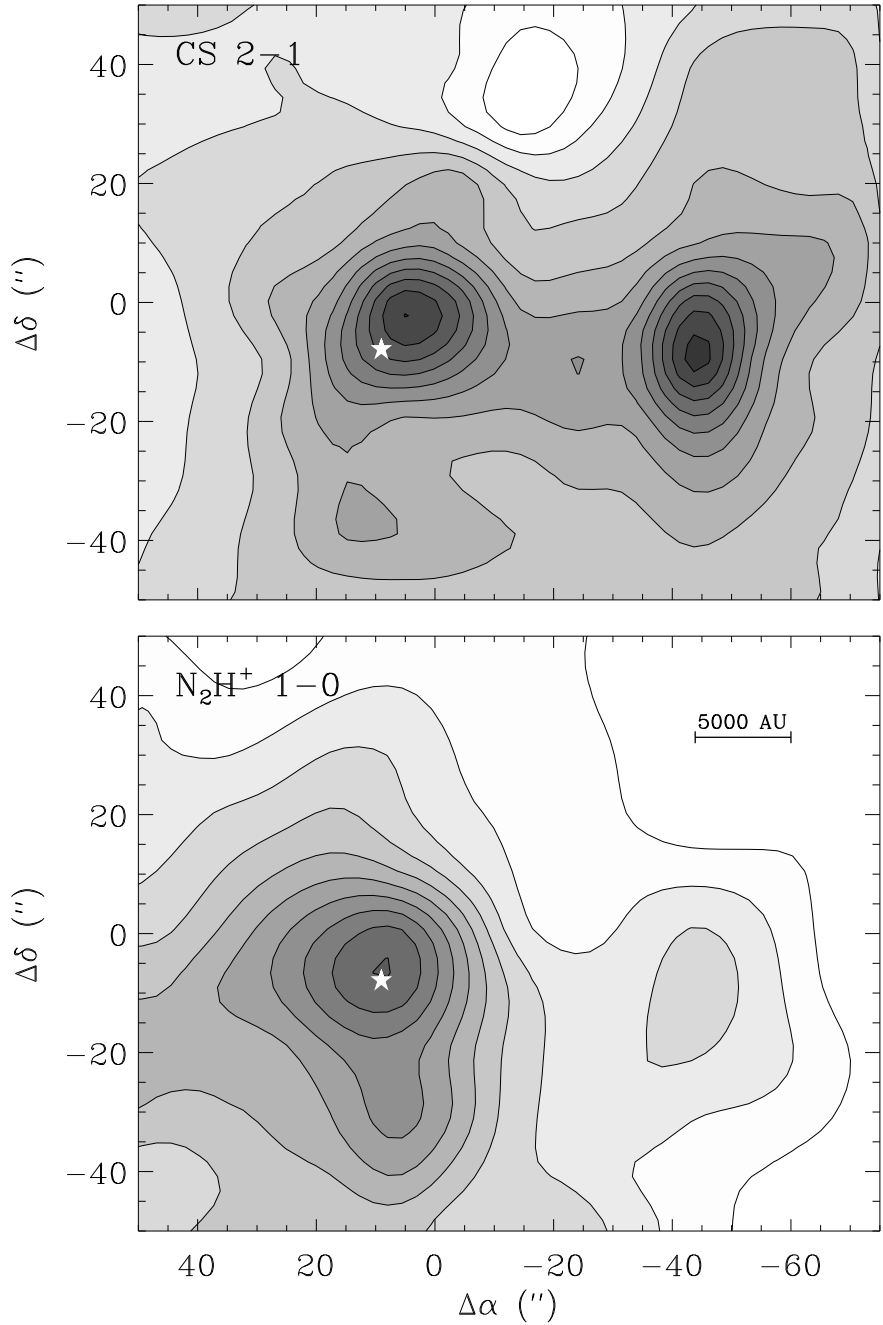,height=9.5in,angle=0,silent=1}}
\vskip -1.0in
\caption{Combined FCRAO/BIMA maps of CS(2--1) and \n2hp(1--0)
of the S68N and S68NW cores. The position of the 3~mm
continuum source (S68N) is marked. Offsets are relative to
$\alpha(2000)=18^{\rm h}29^{\rm m}47\fs 5,
~\delta(2000)=01^\circ 15'51\farcs 4$.
Both maps have been smoothed to a resolution of $10''$.
Contour starting level and increments are 2.0 and 0.5~K~km~s\e\
for the CS and \n2hp\ map respectively.}
\label{fig:maps}
\end{figure}
\clearpage

\begin{figure}[htpb]
\vskip -0.2in
\centerline{\psfig{figure=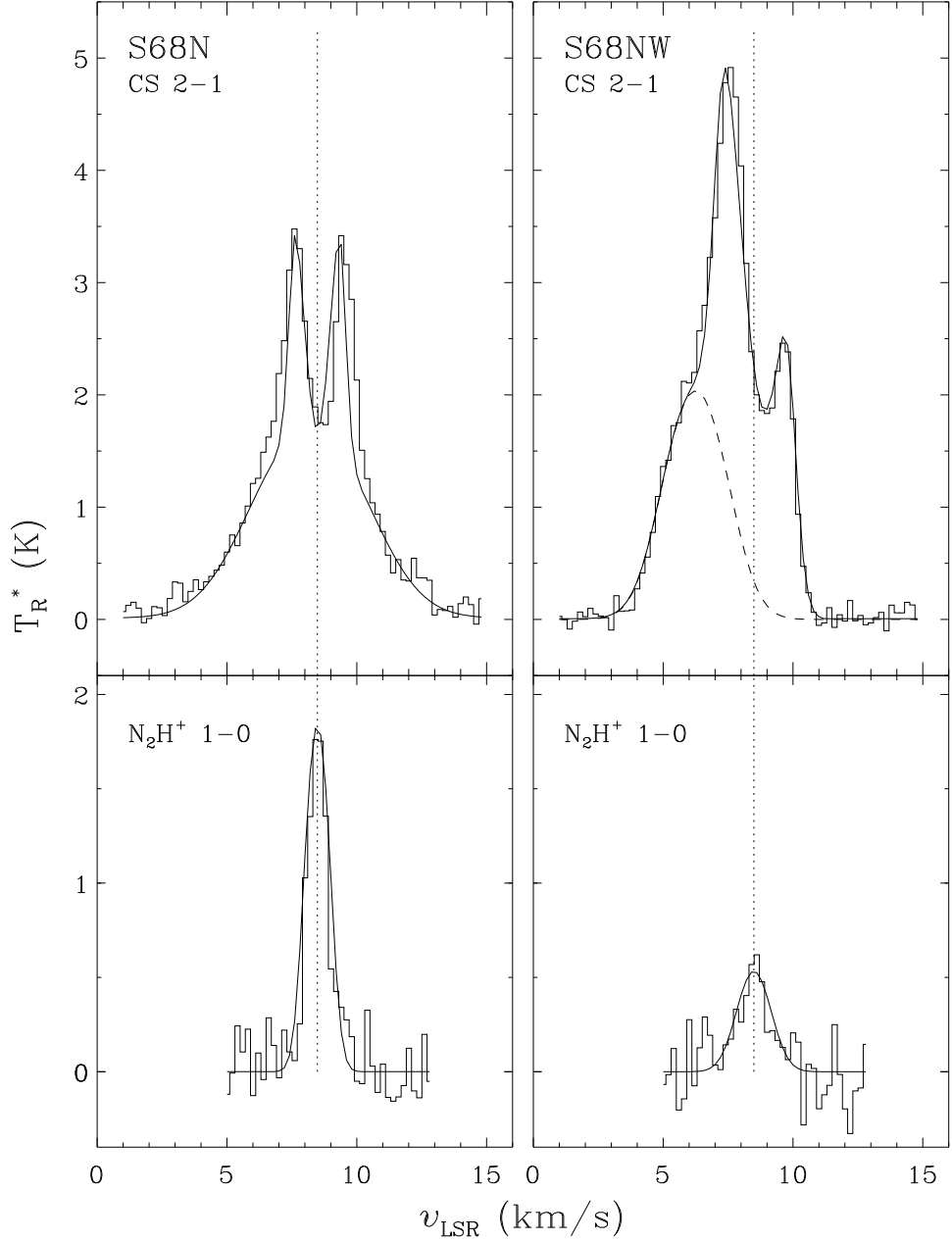,height=7.2in,angle=0,silent=1}}
\vskip 0.0in
\caption{Spectra of the S68N and S68NW cores, averaged over
over the \n2hp\ integrated intensity FWHM contour.
The central velocity of \n2hp\ was measured from a
gaussian fit to the isolated hyperfine component (F$_1$F$=01-12$)
and is indicated by the dotted line. Model fits, consisting of two
layers in relative motion, are shown by the continuous line. An outflow
component has been added to the S68N model to account for the high
velocity wing emission, and an unrelated gaussian component (dashed 
line) to the S68NW model to account for additional emission at low
velocities.}
\label{fig:spectra}
\end{figure}

\end{document}